\documentstyle[preprint,aps]{revtex}
\begin{document}
\draft
\title{Topological Dislocations 
and Mixed State of\\
Charge Density Waves}
\author{Masahiko Hayashi$^{\text{(1),(2),}}$\cite{byline} and 
Hideo Yoshioka$^{\text{(3)}}$}
\address{$^{\text{(1)}}$
Superconductivity Research Laboratory, 
ISTEC, \\ Shinonome 1-10-13, 
Koto-ku, Tokyo 135, Japan}
\address{$^{\text{(2)}}$
Department of Applied Physics, Delft University of Technology,  \\
Lorentzweg 1, 2628 CJ Delft, The Netherlands} 
\address{$^{\text{(3)}}$
Department of Physics, Nagoya University, \\
Furo-cho, Chigusa-ku, Nagoya 
464-01, Japan } 
\date{Received:\ January 24, 1996}
\maketitle

\begin{abstract}
We discuss the possibility of the \lq\lq mixed state\rq\rq \ in 
incommensurate charge density waves with three-dimensional order. 
It is shown that the mixed state can be created by applying an electric 
field perpendicular to the chains. 
This state consists of topological dislocations induced 
by the external field and is therefore similar to the mixed states 
of superfluids (type-II superconductor or liquid Helium II). 
However, the peculiar coupling of charge density waves 
with the electric field strongly modifies the nature of the 
mixed state compared to the conventional superfluids. 
The field and temperature dependence of 
the properties of the mixed state are studied, and 
some experimental aspects are discussed. 
\end{abstract}

\narrowtext

The charge density wave (CDW) is 
an ordered state in which translational 
symmetry is broken \cite{GrunerGorkov}.  
Topological dislocations (TD's)
of incommensurate CDW with 
three-dimensional (3D) order 
have been studied 
by several authors with connection to the 
phase slip phenomena in sliding CDW's 
\cite{Gorkov,OngMaki,Ramakurishuna,Duan}. 
The analogy of the TD's with the dislocations 
in crystals has also been pointed out 
\cite{Feinberg,Bjelis}. 
However, unnoticed up to now seem to 
be the similarities of the TD's with 
the vortices in superfluids: both are topological 
singularities of complex order parameter. 
In type-II superconductors in magnetic fields 
and liquid Helium II in a rotating container, 
the vortices create a remarkable, 
so-called mixed state, which generates 
a wide variety of intriguing phenomena \cite{Vinen}. 
In this Letter, we investigate the possibility of the 
corresponding mixed state in CDW's and discuss its properties. 

In contrast to the vortices in type-II superconductor and liquid Helium 
II, which induce magnetic flux and angular momentum, respectively, 
the TD's in CDW's induce charge polarization. 
Therefore, in analyzing the electrostatic properties of the TD's, 
we have to treat the scalar potential carefully. 
We start with the microscopic model of 
the electron-phonon system in the presence of scalar potential, and 
derive the Ginzburg-Landau free energy of the ordered state. 
The effective free energy of the TD's 
is obtained by integrating out the single-valued part of the phase 
of the order parameter, i.e., the phasons. 
The TD's in the CDW's are similar to the vortices 
in type II superconductors or superfluids, 
except for the coupling to the external field. 
We show that the mixed state is created by an external electric field in 
the direction perpendicular to the chains 
(transverse direction), which can be understood as follows. 
When the system tries to screen the field, the chemical potential 
must be changed from chain to chain 
so as to induce a surface charge \cite{Thorne}. 
This modulation changes the wave number of the CDW condensate, 
i.e., $2 k_{F}$, 
thus conflicts with the 3D order and causes frustration. 
We analyze this state based on the free energy of TD's. 
First, we determined the lower critical field $D_{c1}$ at which 
the first TD appears in the system as we increase the field strength. 
If we increase the field further, the density of the TD's 
also increases. 
In the strong field region, the density of TD's become so high 
that we can treat them as a continuum. 
In this limit we find that the width of the 
mixed state is given by the Thomas-Fermi 
screening length of the normal state. 
If we increase the field even further, the cores of the TD's 
begin to overlap, thus destroying the 3D order. 
The characteristic field of this phenomena, we denote by $D_{c2}$, 
is estimated. 
We argue that these properties should be experimentally 
observable. 

First we derive the free 
energy of CDW from the 1D
electron-phonon system. 
The imaginary-time action reads 
(here we act in the following units $\hbar = k_B =1$ unless noted), 
\begin{eqnarray}
S & = & \int_0^\beta d \tau \bigg[ \int d x 
\psi^*_\sigma (\partial_\tau + i e
\varphi -
\frac{1}{2m}\partial_x^2 - \mu ) \psi_\sigma \nonumber \\ & + & \sum_{q}
b^*_{q}(\tau) (\partial_\tau + \omega_q) b_{q}(\tau) \nonumber \\ & + & 
\frac{1}{\sqrt{L}} \sum_{k,q,\sigma}
g_q a^*_{k+q,\sigma}(\tau)
a_{k,\sigma}(\tau) \left(b_{q}(\tau) + b^*_{-q}(\tau)\right)\bigg],
\label{Sep}
\end{eqnarray}
where $\psi_\sigma = {1}/{\sqrt{L}}
\sum_k
{\rm e}^{i k x} a_{k,\sigma}$ and $b_q$
are variables expressing the electron and phonon degrees 
of freedom with excitation spectra $\epsilon_k = k^2/2m$ and 
$\omega_q$, respectively.
$L$, $g_q$, $\beta$ and $\mu$ are
the length of the chain, the electron-phonon coupling constant, 
the inverse of temperature, and chemical potential, respectively. 
The scalar potential is expressed by
$\varphi$ and the charge of an electron
is given by $-e$. 
From the above action, the free energy of 
the CDW can be derived in
both the cases of $T {\lower -0.3ex \hbox{$<$} \kern -0.75em 
\lower 0.7ex \hbox{$\sim$}} T_c$ and $T \ll T_c$. 
In the former case, we utilize 
the expansion of the order parameter up to
fourth order and the scalar potential up to second order.
On the other hand, in the latter case,
we use a gradient expansion of the
phase of the order parameters and
the scalar potential \cite{Hayashi-Yoshioka-2}.
When the quasi-particle excitation gap $\Delta$ is uniform, 
the free energy becomes in both cases simply, 
\begin{eqnarray}
F & = & \frac{e^2}{\pi v_F}
( 1 - f_s ) \int d x \varphi^2 + i
\frac{e}{\pi} f_s \int d x \theta E 
\nonumber \\ & + & \frac{v_F}{4\pi} f_s 
\int d x (\partial_x \theta)^2 \label{Seffuni}
\end{eqnarray}
disregarding irrelevant constants. 
Here $\theta$ is the phase of the order parameter, $E = - \partial_x 
\varphi$ and $f_s = |\Delta|^2 \pi T \sum_{\epsilon_n}
(|\Delta|^2 + \epsilon_n^2)^{-3/2}$ is the condensate density given by 
$4c_0 |\Delta|^2$ for $T {\lower -0.3ex \hbox{$<$} \kern -0.75em 
\lower 0.7ex \hbox{$\sim$}} T_c$ with $c_0 =
\beta^2 \zeta(3,1/2)/(4 \pi)^2$,
and $1 - \sqrt{2 \pi \beta |\Delta|} 
{\rm e}^{- \beta |\Delta|}$ for $T \ll T_c$, 
where $\zeta(3,1/2)$ is the zeta function. 
The first term of Eq. (\ref{Seffuni}) expresses
the screening due to the excitation of
quasi-particles \cite{VirosztekMaki}.
The second term describes the acceleration of
the CDW and coincides with the result
obtained from the \lq\lq chiral transformation\rq\rq \ 
at $T \ll T_c$
\cite{ShizuyaSakita,NagaosaOshikawa}.

The free energy of the 3D ordered state 
can be obtained by introducing the 
rigidity due to the inter-chain coupling, as follows, 
\widetext
\begin{eqnarray}
F &=& \int d{\bf r} \frac{K}{2} \left[ \left\{\partial_{x} \theta ({\bf r}) 
\right\}^2 + \gamma^2 \left\{\partial_{y} \theta ({\bf r}) 
\right\}^2 + \gamma'^2 \left\{\partial_{z} \theta ({\bf r}) 
\right\}^2 \right] \nonumber \\
&+& \int d{\bf r} \left[ 
i J e \varphi({\bf r}) \partial_x \theta({\bf r}) + 
\frac{1}{8 \pi}
\left\{
|\nabla \varphi ({\bf r})|^2 + \lambda_0^{-2} \varphi ({\bf r})^2 \right\}
+ i e \varphi ({\bf r}) \rho^{ext} ({\bf r}) \right], \label{action1} 
\end{eqnarray}
\narrowtext
where the electrostatic energy is also included. 
We assume that the chains are parallel to the $x$-axis. 
The anisotropy is parameterized by  
$\gamma = \xi_y/\xi_x$ and $\gamma' = \xi_z/\xi_x$ 
with $\xi_x$, $\xi_y$ and $\xi_z$ being
the coherence lengths in $x$-, $y$- and
$z$-direction, respectively. 
The screening length due to the quasi-particles, 
$\lambda_0$, is given by 
$\lambda_0^{-2} = 8 N_\bot e^2 (1 - f_s) / v_F $. 
The response of the system is probed by the external charge 
density $\rho^{ext}({\bf r})$ which induces an external field given by ${\bf 
D}^{ext}_{{\bf k}} = 4 \pi e i {\bf k} \rho^{ext}_{\bf k} / k^2$ where $k^2
\equiv 
|{\bf k}|^2$.
$K = N_\bot v_F f_s/2 \pi$ and $J = N_\bot f_s/ \pi$ 
are coefficients proportional to the areal density of the chains $N_\bot$. 
Note that the size of 
the cores of the TD's is given by $\xi_x$, $\xi_y$ and $\xi_z$ in $x$-,
$y$- and $z$-direction, respectively. 
Although the spatial variation of the 
amplitude of the order parameter is neglected, 
the present treatment is applicable to the most CDW systems 
for the following reason. 
Since the transverse size of the cores 
is usually smaller than the inter-chain spacing 
except near $T_c$, 
the dislocations mostly sit between the chains so 
as to minimize the free energy, 
and the cores thus do not affect the order of CDW. 
In this Letter we focus on the intrinsic 
properties of a clean CDW condensate
and leave the effects of inhomogeneity pinning for further work. 

Based on this free energy, 
we first clarify the screening properties of the CDW state without TD's. 
In this case, only the phason modes contribute to the screening and 
leads the free energy: 
\begin{equation}
F_{eff} = \sum_{\bf k} \biggl[\frac{1}{8 \pi} \left({\bf k}^2 + 
\Lambda_{\bf k}^{-2} \right)
|\varphi_{\bf k}|^2
+ i e \varphi_{\bf k} \rho^{ext}_{-{\bf k}} \biggr],\label{withoutTD} 
\end{equation}
where the effective screening length $\Lambda_{\bf k}$ 
is given as $\Lambda_{\bf k}^{-2} = \lambda_0^{-2} +
(4 \pi e^2 J^2 / K) k_x^2/k_\gamma^2$ and 
$k_\gamma^2 \equiv k_x^2 +
\gamma^2 k_y^2 + \gamma'^2 k_z^2$.
In the low temperature limit, $\lambda_{0}$ diverges 
since the quasi-particle excitations are exponentially 
suppressed by the energy gap, and then 
only the polarization of the condensate can 
contribute to the screening. 
In the direction along the chains (longitudinal direction: 
$k_{y} = k_{z} = 0$), the phason contribution 
completely compensates the suppressed quasi-particle contribution. 
Actually, the screening length becomes 
$(\lambda_0^{-2} + 4 \pi e^2 J^2 / K)^{-1/2}$, 
which coincides with the Thomas-Fermi
length of the normal state, $\lambda_{TF} \equiv
\{ 8 N_\bot e^2/(\hbar v_F) \}^{-1/2}$. 
On the other hand, in the transverse direction ($k_{x} = 0$), 
there is no phason contribution and, consequently, no screening 
for $T \ll T_c$. 
This can be attributed to the  \lq\lq rigidity\rq\rq \ of CDW. 

Next we examine how the TD's affect the screening 
properties. 
Here we limit our discussion to 
straight dislocations parallel to the $z$-axis, 
which reduces the problem to a two-dimensional one. 
The density of topological charge is then 
$n({\bf r}) = \sum_\mu q_\mu \delta^{(2)}
({\bf r} - {\bf r}_\mu)$, where $q_\mu$ and
${\bf r}_\mu$ are the topological charge, 
$q_\mu = 2 \pi \times \text{(integer)} $, 
and the position of the $\mu$-th dislocation, respectively. 
The phase of the order parameter is given by the 
relation,  
$(\nabla \theta)_{\bf k} = 
i {\bf k} \theta^s_{\bf k} + 
i {\bf k} \times {\hat{\bf z}} \ n_{\bf k} / k^2$, 
where $\theta^s_{\bf k}$ expresses the phason part, 
${\hat{\bf z}} = (0,0,1)$, and 
${\bf k} = (k_{x},k_{y},0)$. 
In the following  we consider TD's with $q_\mu = \pm 2 \pi$ only. 
By integrating out the phason mode,
the free energy of TD's per unit length becomes, 
\widetext
\begin{eqnarray}
\frac{F_{eff}}{L_z} = \sum_{\bf k}
\biggl[&& \frac{K \gamma^2}{2} \frac{1}
{k_\gamma^2} \left\{n_{\bf k} - \frac{i e
J}{K} \left(i k_y \varphi_{\bf k}\right)
\right\} \left\{n_{-{\bf k}} - \frac{i e
J}{K} \left( - i k_y \varphi_{-{\bf k}}
\right) \right\}\nonumber\\
&&+ \frac{1}{8 \pi} \left(k^2 +
\Lambda_{TF}^{-2}\right)
|\varphi_{\bf k}|^2 + i e \varphi_{\bf k}
\rho^{ext}_{-{\bf k}} \biggr],
\label{effectiveaction2}
\end{eqnarray}
\narrowtext
\noindent
where $L_z$ is the length of the sample
in the $z$-direction.
Note that in the imaginary time formulation 
the scalar potential has to be rotated to the imaginary axis like
$\varphi_{\bf k} \rightarrow i
\varphi_{\bf k}$. 
Therefore we define the expectation value of $\varphi_{\bf k}$ by
$\langle \varphi_{\bf k} \rangle \equiv i
{\bar \varphi}_{\bf k}$, where ${\bar \varphi}_{\bf k}$ is 
the physical value. 

Eq. (\ref{effectiveaction2}) tells us that the TD's behave like a 
Coulomb gas with background charge $- e J (k_y \varphi_{\bf k})/K$, 
which is proportional to the electric field in $y$-direction. 
Therefore TD's are generated by the external 
electric field in $y$-direction, 
forming the \lq\lq mixed state\rq\rq \  of CDW. 
In order to quantify this point, we study two limiting cases: 
the dilute and the dense limit of dislocations. 

In the dilute limit, we consider the problem of a single TD: 
When does the first TD appear as we increase the field strength? 
The corresponding critical field can be estimated from the difference of 
the free energy between the single TD state and 
the TD free state, $\Delta F$. 
It is given in the limits of $T \ll T_{c}$ and 
$T {\lower -0.3ex \hbox{$<$} \kern -0.75em 
\lower 0.7ex \hbox{$\sim$}} T_c$ as, 
\begin{equation}
\frac{\Delta F }{L_{z}} = 
\left\{
\begin{array}{ll}
\pi K \gamma W \lambda_{TF}^{-1}
- 2 \pi e J D^{ext}_{y} W^2 &\text{ for $T \ll T_{c}$,}
\\
\pi K \gamma \ln (W / \xi_{y}) 
- 2 \pi e J \lambda_{TF}^2 D^{ext}_{y}&\text{ for 
$T {\lower -0.3ex \hbox{$<$} \kern -0.75em 
\lower 0.7ex \hbox{$\sim$}} T_c$.}
\end{array}\right. 
\label{deltaF}
\end{equation}
In calculating Eq.(\ref{deltaF}), 
we assumed that the system is infinite in the $x$-direction 
and of width $W$ in the $y$-direction, 
thus introducing an infrared cut-off of $1/W$ for the wave vector 
$k_{y}$. 
Note that \lq\lq $T {\lower -0.3ex \hbox{$<$} \kern -0.75em 
\lower 0.7ex \hbox{$\sim$}} T_c$\rq\rq \  should not include 
the region too close to $T_{c}$, 
where the present treatment fails due to the diverging 
coherence length. 
In the limit of $T \ll T_{c}$, 
the critical field is expressed as 
$D_{c1} = (\gamma / 4 e ) \hbar \omega_{p} \times 
W^{-1}$, where $\omega_{p}$ is the plasma frequency 
given by $\sqrt{8 N_{\bot} e^2 v_{F} / \hbar}$. 
Therefore, the critical voltage defined by $V_{c1} \equiv D_{c1} W$ 
becomes independent of the sample width. 
On the other hand, at $T {\lower -0.3ex \hbox{$<$} \kern -0.75em 
\lower 0.7ex \hbox{$\sim$}} T_c$,  
$D_{c1}$ is given by $(\gamma / 4 e)  \hbar \omega_{p}
\times \lambda_{TF}^{-1} \ln (W/\lambda_{TF})$. 
This is larger than the $D_{c1}$ of the low temperature limit, 
because the energy gain due to 
the creation of TD's is reduced by quasi-particles. 

In the dense limit, $n_{\bf k}$ is approximated 
by a continuous function rather than a function of 
the positions of individual TD's $\{{\bf r}_\mu\}$. 
This corresponds to neglecting the spatial structure
smaller than the spacing between TD's.
In this approximation, 
it is quite simple to evaluate $n_{\bf k}$ 
and $\bar \varphi_{\bf k}$ by using 
Eq. (\ref{effectiveaction2}).
The minimization of the free energy $F_{eff}$ in terms of $n_{\bf k}$ 
gives,
\begin{equation}
n_{\bf k} = - \frac{e J}{K} k_y \varphi_{\bf k}.
\label{tddensity}
\end{equation}
The density of TD's turns out to be proportional to the strength of the
electric field in
the $y$-direction.
On the other hand, the variation of 
$F_{eff}$ with respect to $\varphi_{\bf k}$ gives the Poisson equation, 
\begin{equation}
(k^2 + \Lambda_{\bf k}^{-2}) {\bar \varphi_{\bf k}} =
- 4 \pi e \left(
J \frac{\gamma^2}{k_\gamma^2}i k_y n_{\bf k} + \rho_{\bf k}^{ext}
\right). 
\label{poisson}
\end{equation}
Substituting Eq. (\ref{tddensity}) into Eq. (\ref{poisson}), 
we obtain, ${\bar \varphi}_{\bf k} = 
- 4 \pi e\rho^{ext}_{\bf k} /(k^2 + \Lambda_m^{-2})$, 
where $\Lambda_{m}^{-2} = \lambda_{0}^{-2} + 4 \pi e^2 J^2 / K$. 
Note that $\Lambda_m$ equals now the 
Thomas-Fermi screening length of the normal state, 
$\lambda_{TF}$. 
When we apply the electric field in 
the $y$-direction, the field penetrates the
sample to the depth $\Lambda_m$.  
As we can see from Eq. (\ref{tddensity}), the TD's are 
thereby distributed in this surface region where 
the mixed state is formed. 

In case of type-II superconductors 
(or liquid Helium II), 
there is another characteristic field, i.e., the upper 
critical field (or angular velocity), usually denoted as 
$H_{c2}$ ($\Omega_{c2}$) \cite{Vinen}, which defines a phase transition 
within the mean field theory. 
$H_{c2}$ ($\Omega_{c2}$) can be understood as the field at which the 
cores of the vortices begin to overlap. 
In case of the CDW's, when the cores of the TD's begin to overlap, 
the inter-chain ordering is almost destroyed 
by strong fluctuations due to the dimensional reduction. 
We denote this characteristic field as $D_{c2}$ for convenience, 
although it is still unclear whether 
there is a real phase transition or not. 
This field strength is first realized at the surface where the 
internal field coincides with the external one. 
Since the cores begin to overlap when the number density of the 
TD's exceeds $1/(\xi_{x}\xi_{y})$, 
the corresponding field strength, $D_{c2}$, is estimated from Eq.
(\ref{tddensity}) 
as $D_{c2} = \pi v_{F} \hbar/(e \xi_{x}\xi_{y})$. 

In the following, we discuss the experimental aspects 
of the mixed state with actual CDW materials in mind. 
We concentrate on the low temperature limit where the 
effects due to the mixed state are rather clear. 
Since the TD's modify the screening properties of the system, 
the mixed state can be observed by measuring the dielectric constant of the
system in a capacitor with the transverse electric field. 
The capacitance should display the crossover from the insulating behavior in 
the weak field region ($V < V_{c1}$) to the metallic one in the strong
field region ($V > V_{c1}$). 
In the case of $K_{0.3} Mo O_{3}$ ($\gamma = 0.01 \sim 0.1$ 
depending on the direction\cite{GrunerGorkov}) , 
$V_{c1}$ is estimated as $\gamma \times 0.69$ V. 
On the other hand, $D_{c2}$ is estimated as 
$\gamma^{-1} \times 2.9 \times 10^{6}$ 
V/m, which can be much larger than $V_{c1}$ even if we take 
$W = 1$ $\mu \text{m}$. 
It should be noted that, in the case of $K_{0.3} Mo O_{3}$, 
a simple estimation of the Thomas-Fermi screening length 
in the normal state, $\lambda_{TF} = \omega_{p}/v_{F}$, 
gives less than 1 ${\text{\AA}}$, 
which is smaller than the inter-chain spacing $\sim 10$ 
${\text{\AA}}$. 
This estimation is based on the continuum model of metals and, 
therefore, we cannot take it too seriously since the continuum 
approximation is no longer valid at this scale, 
especially in the transverse direction. 
However, we expect that, in the strong field region, 
the width of the mixed state can be as narrow as the 
lattice spacing and most TD's are confined to the outermost layers. 
In the weaker field region, this width can be much wider. 

In the uniform electric field, there can be an additional 
effect due to the quasi-particles excited 
through Zener tunneling process \cite{Monceau}. 
In the WKB approximation \cite{Zener}, 
the characteristic electric field of 
Zener breaking is huge in the present configuration 
($\sim \gamma^{-1} \times 10^7$ V/m)
due to the narrow band width and the large lattice spacing. 
Therefore the contributions from the quasi-particle 
excitations is strongly suppressed and we expect 
that the mixed state of the condensate can be observed. 

The effects of pinning cannot be avoided in realistic CDW systems
\cite{FLR}, but are beyond the scope of the present Letter. 
However, we expect a hysteretic behavior in the observed capacitance. 
Of course, the magnitude of the hysteresis depends strongly 
on the inhomogeneity of the sample. 

In conclusion, we proposed a new state, i.e., the mixed state, 
in CDW's, which is realized by applying the transverse electric field. 
The physical properties are quantitatively discussed 
based on the Ginzburg-Landau free energy of the CDW and 
the critical fields which characterize 
the mixed state are estimated. 
The mixed state strongly affects 
the screening properties of the system and is thus detectable 
through the dielectric constant measurement in a capacitor. 

The authors wishes to thank Prof. H. Fukuyama and 
Prof. H. Matsukawa for stimulating discussions 
and, Prof. G. Bauer, Dr. Y. Nazarov and Dr. C. Dekker 
for useful comments.

\end{document}